\documentstyle[12pt,aasms4]{article}

\lefthead{}
\righthead{}

\def\asec {^{\prime\prime}}
\def\amin {^{\prime}}
\def\fsec  {{\rlap.}^{\rm s}}
\def\fasec {{\rlap.}^{\prime \prime}\hskip0.05em}
\def\deg  {^\circ}

\begin{document}

\title{DISCOVERY OF A NUCLEAR X-RAY SOURCE IN NGC 7331:\\
EVIDENCE FOR A MASSIVE BLACK HOLE}

\author{Christopher J. Stockdale,
W. Romanishin,
and John J. Cowan}

\affil{Department of Physics and Astronomy, 
University of Oklahoma, Norman, OK 73019}

\begin {center}
stockdal@mail.nhn.ou.edu,wjr@mail.nhn.ou.edu,cowan@mail.nhn.ou.edu
\end {center}

\begin{abstract}
Using ROSAT, we have made the first deep X-ray observations of NGC~7331, discovering a
nuclear X-ray source coincident with the previously determined radio and optical nuclear centers.  Positions, luminosities, and fluxes of X-ray sources in the field are compared
to previously identified radio and optical sources.
The nucleus of NGC~7331 has been analyzed to discern any new
evidence supporting the presence of a massive black hole (MBH), and comparisons
are made with other Low-Ionization Nuclear Emission-line Region (LINER)
galaxies.  The flux ratio of core radio to soft X-ray emission in NGC~7331
is lower than in other LINERs included in our
sample.
\end{abstract}


\section{Introduction}

Deep radio observations of NGC~7331 were first made in an attempt to
search for the supernova remnant (SNR) from SN1959D (Cowan, Romanishin, \& Branch 1994, hereafter = CRB).  Although the SNR was
not identified in those observations, an unresolved nuclear source was detected (CRB).  
Optical observations of the nucleus of NGC~7331
identified an optical LINER spectrum (Keel 1983; Bower 1992).  These two
discoveries support the possibility that NGC~7331 harbors an MBH.  
Rubin et al. (1965) and Afanasiev, Silchenko, \& Zasov
(1989) made kinematic studies, measuring
rotational motion from [NII] and H$\alpha$ emission lines.  
Based on these observations were analyzed by Afanasiev et al.(1989)
 suggested an MBH of $\sim$ $5\times10^{8}$ $M_\odot$ in NGC~7331.  
Bower et al. (1993), however, suggested that this rotational motion could be explained by models without an MBH 
and with a constant mass-to-light ratio as a function of radius.
Bower et 
al. (1993) were only able to exclude an MBH with a mass greater than 
$5-10\times10^{8}$ $M_\odot$ (CRB), a result which is relatively insensitive to small differences in the assumed distance to the galaxy. 
The Bower et al. (1993) mass limit is also 
near the upper limit of measured masses for Massive Dark Objects found in LINER 
galaxies, which have X-ray and radio properties similar to those found in 
NGC~7331 (Ho 1998).

In this Letter we report on the first deep ROSAT X-ray observations of NGC~7331
and their impact in understanding the nature of its LINER nucleus.   
In \S 2, we discuss the
results of the observations and the techniques used to reduce these data.
Classically, the possible indicators for an MBH are the detection of an 
unresolved,
nuclear source, a non-thermal radio and X-ray spectral index, variability
in the nuclear flux over months to years, and an anti-correlation between X-ray
and radio variability (Melia 1992).  The fact that NGC~7331 possesses an 
unresolved nuclear source and has a non-thermal radio spectral index was 
confirmed by CRB.  This Letter identifies a similarly unresolved X-ray source, 
but can make no statement as to the nature of the X-ray spectra.  Confirmation 
of a non-thermal X-ray spectra and the last
two criteria for an MBH require further X-ray and radio observations.  While starburst galaxies can also have non-thermal, unresolved sources, there is no evidence for an anti-correlation in X-ray and radio variability in such galaxies.  Other indicators of the presence of an MBH in NGC~7331 will be discussed later in \S 3 of this Letter. Also in \S 3, we will provide comparisons of our results with current theories and observations of low luminosity AGN (LLAGN) powered LINERs and make our conclusions.

\section{Observations and Data Reduction}

NGC 7331 is an Sbc spiral galaxy, which was observed with the ROSAT High
Resolution Imager (HRI) in December 1995 to study the nuclear source that
had been previously observed in the radio with the Very Large Array (VLA) (CRB).
The HRI is an X-ray photon collector with no spectral 
resolution, collecting photons with energies ranging from 0.12 keV to 2.4 keV.  
The observations were made over a two day period with a total observation time of 
30.5 ks.  The search was centered on the CRB radio center of NGC~7331 
at R.A.(2000) $ = 22^{h}37^{m}4\fsec 8$, 
Decl.(2000)$ = +34\deg 25\amin 11\fasec 5$.

Raw data reduction was provided by the ROSAT Standard Data Processing Center.
Standard ROSAT tests for short term variability are not applicable due to
insufficient counts.  The data were further reduced using the IRAF
routine GAUSS which convolved the data with a circularly symmetric Gaussian
function, with the parameter sigma set to one pixel, effectively $8\fasec 0$.  The image was smoothed with a Gaussian because of concerns from the Data Processing Center that the standard analysis system was not very reliable in properly identifying sources in complex emission regions.  Sources that were initially reported as individual point sources by the Center, are actually part of a more complex X-ray emission region.  The data were then evaluated with the NOAO-IRAF routine QPHOT to obtain
object counts, count noise, and accurate position measurements.  The aperture radius used to determine the count flux from the galaxy was set to $24\arcsec$, and sky background measurements were determined with the specified
parameters of an inner annulus of $80\arcsec$ and outer annulus of $102\arcsec$.  

Lacking direct X-ray spectral data, a value for the $ log N(H)=21.3$ (Burstein \& 
Heiles 1978; van Steenberg et al. 1988; Stark et al. 1992; Smith 1998) and a photon index, $\Gamma$
 ($dF/dE \propto E^{-\Gamma}$), of 1.0 and 2.0 for the 
central sources in NGC~7331 were assumed to determine the unabsorbed X-ray flux. The total column density of hydrogen includes both
the Galactic column density (Burstein \& Heiles 1978) and the column density for the bulge of NGC~7331 (Smith 1998).  The ROSAT HRI energy-to-count conversion factors were taken from the 
ROSAT Users Handbook, assuming a power law spectrum.
The assumption of an X-ray power law spectrum is consistent with both
the possible presence of an MBH, as well as for a post starburst galaxy (Fabbiano 1996).
The X-ray luminosity for the galaxy was calculated assuming the distance to NGC~7331 as 15.1 Mpc 
(Hughes et al. 1998).  The detections reported in Table 1 all meet at least a 3$\sigma$ detection threshold.  Since little is known about the other detected sources in Table 1, a power law spectrum, a photon index of 1.0 and the same line-of-sight, Galactic column density, which was used for NGC~7331, were used to determine their energy flux measurements.

Sources of uncertainty in ROSAT positions include a known attitude
solution error, which causes X-ray position offsets of order $6\asec$ 
(ROSAT Users Guide).  Since any correction would require knowing X-ray
positions and this is the first high resolution image of the field, it
is not possible to accurately correct for this error.  Table 3 compares
the optical, radio, and X-ray positions of the nucleus of NGC~7331.  The
values for the nuclear positions agree to well within $5\arcsec$ and are definitely coincident within the acceptable ROSAT error.
 
\section{Results and Discussion}

The observations of the X-ray nuclear source in NGC~7331 support the
likelihood of an MBH in its core.  In Table 2, we compare the luminosity of
the nuclear source in NGC~7331 with
luminosity values for other LINER galaxies. Also included in this table for comparison are the
reported luminosities for two, identified starburst galaxies.

An assumed power-law spectrum and photon indices of 1.0 and 2.0 were used to analyze our ROSAT observations of NGC~7331, because
these are typical of other observed X-ray LINERs in the
energy range from 0.1 keV to 2.4 keV.  
The X-ray luminosity of NGC~7331 in this
bandpass region is within the observed range found in other LINERs and somewhat larger than those luminosities observed in typical starburst galaxies.  We note 
that these LINER X-ray luminosities are near the upper limit of X-ray luminosities typically found in normal spiral galaxies (Fabbiano 1996), and therefore an X-ray luminosity cannot be used solely to identify possible MBHs.

The radio spectrum of a typical LLAGN is predicted to be powered by
cylcosynchrotron emissions in the advection dominated accretion flow (ADAF) model used by Mahadevan (1997).   The observed spectral index, $\alpha$, of $-0.6$ ($S_{\nu} \propto \nu^{+\alpha}$) for NGC~7331 indicates a non-thermal
source powering the radio spectrum between 6 cm and 20 cm and is in agreement with
observations of similar sources listed in Table 4.  The value reported by CRB for the spectral index of NGC~7331 agrees closely with the theoretical value of the spectral index of M31, which was modeled with an MBH by Melia (1992).  It should also be noted that this value of $\alpha$ is only slightly greater than the spectral index normally associated with SNRs, $-0.8$,  and this might suggest a post starburst nature (Condon 1996).

Table 4 also lists radio fluxes from other LINERs with identified nuclear radio sources.  The same value of the spectral index found for NGC~7331 is reported for M51 and M81 by Turner \& Ho (1994), and M81 has been confirmed to harbor an LLAGN by Ho et al. (1996).    While the CRB spectral index alone does not confirm the exact nature of the nuclear source in NGC~7331, it does suggest that NGC~7331 is an LLAGN, given the galaxy's identification as a LINER galaxy and the similarity of its spectral index with other LINER galaxies confirmed to harbor MBHs.

For a sample of LINER nuclei Table 5 lists the ratio of soft X-ray and 6 cm radio fluxes.  The X-ray fluxes listed in Table 5 are averaged over their stated bandpasses listed in Table 2.  Among this limited sample, there appears to be a range for galaxies, from being relatively radio quiet (or X-ray loud)  (log [X-Ray/Radio] $\propto -1.8$) to relatively radio loud (or X-ray quiet) (log [X-Ray/Radio] $\propto -4.6$). It must be noted that this is a small sample and that these observations were taken at various epochs and with different instruments.  (Radio observations are from the VLA and the Westerbork Synthesis Radio Telescope and X-ray observations are from ROSAT and ASCA.)

Another predicted indicator of an MBH is the bipolar
outflow, which is believed to develop as the result of convective instabilities
in the thin disk approximation.  This in turn leads to a quasi-spherical accretion flow,
characteristic of the advective model (Narayan \& Yi 1995).  In the case of 
NGC~7331, these outflows have not been positively identified in any observations to date.  If these features do exist, they are likely to extend only a few
parsecs.  For NGC~3079, a LINER galaxy 16 Mpc away, jets were identified which
extended only 1.5 pc from the central engine (Trotter et al. 1998).  Trotter et
al. (1998) detected a molecular disk with a binding mass of $\sim$~$10^{6}$~$M_\odot$.
The radio observations of NGC~7331 indicate only a
compact nuclear source with a ring of non-thermal sources extending beyond
the optical galaxy (CRB).
The fact that similar jets have not been detected in NGC~7331 can be attributed
to the galaxy being 15.1 Mpc away ($1\asec$ = 73 pc), which makes resolution of parsec-scale outflows beyond the capability of the CRB observations.

Further evidence supporting the existence of a LINER nucleus in NGC~7331, 
similar to the one in M31, has been presented by Mediavilla et al. (1997) and further confirmed by Heckman (1996).  M31 was identified as harboring an MBH of $3$ $\times$ $10^{7}$ $M_{\odot}$ by Kormendy \& Richstone (1995).  Mediavilla 
et al. (1997) and Ciardullo et al. (1988)
report that the kinematics of the stars and ionized gas are decoupled, in
 NGC~7331 and M31 respectively.  This is
in contrast to the conditions found in Seyfert galaxies.  Also, Tosaki 
\& Shioya (1997) and Young \& Scoville (1982)
found no CO emission near the nuclei of NGC~7331 and M31.  This is 
atypical for post starburst galaxies.  Their results do support
the possibility that this absence of CO emission may be due to the presence of
an MBH. 

Also identified in the HRI image was a source which is within $9\asec$ of
the optical position of NGC~7335 (Klemola et al. 1987).  The observed flux
was $2.15\times 10^{-13}$ erg sec$^{-1}$ cm$^{-2}$
(assuming a power-law spectrum, a photon index of 1.0, and the same line
of sight, Galactic column density of hydrogen used in the previous section).
If this
X-ray source is associated with NGC~7335, this would indicate an X-ray luminosity of $3.1\times 10^{41}$ erg s$^{-1}$, assuming a distance to NGC~7335 of 110 Mpc (z=6315 km/s de Vaucouleurs et al. (1991) and
a Hubble Constant of 55 km s$^{-1}$ Mpc$^{-1}$).  Few observations
have been made of this galaxy, and little is known about it.  Based upon this derived X-ray
luminosity, however, this galaxy warrants further study to determine if it is also a possible candidate for AGN activity.

 There are a number of tests associated with X-ray studies that could more definitively determine the nature of the nucleus of NGC~7331 (see e.g., Fabbiano 1996; Serlemitsos et al. 1996; Nandra et al. 1997; Ho 1998).  Such tests will require deep observations, which could provide resolution of the
spectrum at multiple wavelengths and identify any form of variability in the observed flux from NGC~7331.

\acknowledgments

The research was supported in part by NASA Grant NAG5-3214 and has made use of the NASA/IPAC Extragalactic Database (NED), which is operated by the Jet Propulsion Laboratory, Caltech, under contract with the National Aeronautics and Space Administration.  We thank an anonymous referee for helping us to improve this Letter.

\clearpage

\begin{table}

\begin{tabular}{cccccc}
\multicolumn{6}{c}{TABLE 1.  X-Ray Sources in the Field of NGC~7331} \vspace{0.2cm} \\ \hline\hline
\multicolumn{1}{c}{Source} &\multicolumn{1}{c}{RA} &\multicolumn{1}{c}{Decl.}
&\multicolumn{1}{c}{Counts} &\multicolumn{1}{c}{Flux} &\multicolumn{1}{c}{Comments}\\
  & (2000) & (2000) & (ksec$^{-1}$) & ($10^{-13}$ erg s$^{-1}$ cm$^{-2}$) & \\ \hline
1 & $22^{h}37^{m}3\fsec 64$ & $+34\deg 24\amin 59\fasec 27$ & $4.17\pm .4$ & $41.3\pm .63$ & NGC~7331\\
2 & $22^{h}37^{m}1\fsec 86$ & $+34\deg 26\amin 53\fasec 90$ & $1.3\pm .3$ & $2.15\pm .36$ & NGC~7335\\
3 & $22^{h}36^{m}29\fsec 38$ & $+34\deg 25\amin 56\fasec 78$ & $0.9\pm .3$ & $1.49\pm .50$ &\\
4 & $22^{h}36^{m}33\fsec 13$ & $+34\deg 30\amin 11\fasec 49$ & $1.9\pm .4$ & $3.16\pm .66$ & \\
5 & $22^{h}37^{m}28\fsec 97$ & $+34\deg 25\amin 25\fasec 49$ & $1.0\pm .3$ & $1.66\pm .50$ & \\
6 & $22^{h}37^{m}38\fsec 15$ & $+34\deg 22\amin 27\fasec 36$ & $0.5\pm .3$ & $0.83\pm .50$ &\\
7 & $22^{h}37^{m}3\fsec 48$ & $+34\deg 9\amin 47\fasec 85$ & $1.4\pm .7$ & $2.33\pm 1.16$ &\\
8 & $22^{h}37^{m}51\fsec 93$ & $+34\deg 28\amin 29\fasec 43$ & $0.9\pm .4$ & $1.49\pm .66$ &\\
9 & $22^{h}35^{m}54\fsec 60$ & $+34\deg 13\amin 51\fasec 34$ & $9.2\pm .9$ & $15.3\pm 1.5$ &\\
10 & $22^{h}36^{m}3\fsec 43$ & $+34\deg 12\amin 14\fasec 95$ & $5.9\pm .9$ & $9.80\pm 1.50$ &\\

\hline

\end{tabular}
\end{table}

\clearpage
\begin{table}

\begin{tabular}{ccccccc} 
\multicolumn{7}{c}{TABLE 2. Comparisons of NGC 7331 and} \\
\multicolumn{7}{c}{Other Similar X-ray Sources} \vspace{0.2cm} \\ \hline\hline
\multicolumn{1}{c}{Source}   &\multicolumn{1}{c}{Type}
&\multicolumn{1}{c}{Luminosity} &\multicolumn{1}{c}{Photon Index} &\multicolumn{1}{c}{Bandpass} &\multicolumn{1}{c}{Mechanism} &\multicolumn{1}{c}{Ref}\\

 	 & (RC3) & $10^{40}$ erg s$^{-1}$  & & keV & &  \\ \hline

NGC 7331 & SAab   & $2.70$ --- $8.46$ & $1.0<\Gamma <2.0$ & 0.1-2.4 & LINER & this Letter\\  
NGC 4258 & SABbc & $0.51$       & $3.7$           & 0.1-2.4 & LINER & 1\\
NGC 4736 & R SA(r)ab & $0.34$       & $2.3$           & 0.1-2.0 & LINER & 2\\
NGC 4594 & SAa	 & $5.3$	& $1.80$	  & 0.5-2.0 & LINER & 3\\
M51	 & SAbc	 & $2.5$	& $1.76$	  & 0.5-2.0 & LINER & 3\\
NGC 3079 & SBc	 & $4.4$	& $1.76$	  & 0.5-2.0 & LINER & 3\\
M81	 & SAab	 & $1.2$	& $1.9$		  & 0.5-2.0 & LINER & 3\\
NGC 3642 & SA(r)bc & $3.9$	& $2.84$	  & 0.2-2.2 & LINER & 4\\
NGC 3147 & SA(r)bc & $24$	& $1.74$	  & 0.5-2.0 & LINER & 3\\
NGC 4579 & SAB(rs)b & $28$ 	& $1.87$	  & 0.5-2.0 & LINER & 3\\
M33	 & SAcd    & $0.1248$     & $0.74$          & 0.1-2.4 & Starburst  & 5\\
NGC 3628 & SAb    & $<0.06$---$1.7$ & $0.27$---$1.16$  & 0.1-2.0 & Starburst  & 6\\

\hline
\end{tabular}\newline
References \newline
1.  Pietsch et al. (1994) \newline
2.  Cui et al. (1997)    \newline
3.  Serlemitsos et al. (1996) \newline
4.  Koratkar et al. (1995) \newline
5.  Long et al. (1996)   \newline
6.  Dahlem et al. (1995) \newline
\end{table}

\clearpage
\begin{table}
\begin{tabular}{cllc}
\multicolumn{4}{c}{TABLE 3.  Nuclear Positions for NGC 7331} \vspace{0.2cm} \\
\hline\hline

Observation & R.A. (2000) & Decl.(2000) & Reference
\vspace{0.2cm} \\ \hline

X-ray & $22^{h}37^{m}3\fsec 64$   
& $+34\deg 24\amin 59\fasec 27$     & this Letter \\
Radio & $22^{h}37^{m}4\fsec 00$   
& $+34\deg 24\amin 57\fasec 07$     & CRB (1994)\\
Optical & $22^{h}37^{m}4\fsec 01$   
& $+34\deg 24\amin 56\fasec 07$     & Klemola 1994, private communication\\
Optical & $22^{h}37^{m}4\fsec 04$   
& $+34\deg 24\amin 56\fasec 47$     & Argyle \& Clements (1990) \\
\hline\hline
 & & &  \\
\end{tabular}
\end{table}
\clearpage
\begin{table}

\begin{tabular}{cccccccc} 
\multicolumn{7}{c}{TABLE 4. Core Flux Comparisons of NGC 7331 and} \\
\multicolumn{7}{c}{Other Similar Radio Sources} \vspace{0.2cm} \\ \hline\hline

\multicolumn{1}{c}{Source}   &\multicolumn{1}{c}{2cm} 
&\multicolumn{1}{c}{3.6cm} &\multicolumn{1}{c}{6cm} &\multicolumn{1}{c}{20cm} &\multicolumn{1}{c}{Spectral Index} & \multicolumn{1}{c}{Ref}\\

Type 	 & mJy & $\mu$Jy & mJy & mJy & & \\ \hline 
NGC 7331 & $\cdots$ & $\cdots$ & $0.121$ & $0.234$ & $-0.6$ & 1\\
SAab      & $\cdots$ & $\cdots$ & (B array) & (A array) & & \\ \hline
NCG 4258 & $3.2$ & $\cdots$ & $2.4$ & $<10$ & $+0.4$ & 5 \& 2\\
         & $\cdots$ & $\cdots$ & $1.5$ & $2.0$ & $-0.2$ & 3\\    
SABbc	 & (C array) & $\cdots$ & (B array) & (see ref) \& (A array) & & \\ \hline
NCG 4579 & $\cdots$ & $\cdots$ & $\cdots$ & $40$ & $\cdots$ & 2\\ 
SAB(rs)b & $\cdots$ & $\cdots$ & $\cdots$ & (see ref) & & \\ \hline
NCG 4736 & $3.2$ & $\cdots$ & $6.1$ & $27$ & $-0.3$ & 2 \& 5\\ 
         & $2.2$ & $\cdots$ & $3.6$ & $\cdots$ & $-0.4$ & 5\\
R SA(r)ab & (C array) & $\cdots$ & (B array) & (see ref) & & \\ \hline
M31     & $\cdots$ & $28$ \& $39$ & $\cdots$ & $\cdots$ & $\cdots$ & 4\\
SA(s)b   & $\cdots$ & (A array) & $\cdots$ & $\cdots$ & &\\ \hline
M51     & $2.3$ &  $\cdots$ & $1.6$ & $\cdots$ & $-0.6$ & 5\\
SAbc     & (C array) & $\cdots$ & (B array) & $\cdots$ & & \\ \hline
M81     & $45$ &  $\cdots$ & $92$ & $\cdots$ & $-0.6$ & 5\\
SAbc     & (C array) & $\cdots$ & (B array) & $\cdots$ & & \\ \hline
 
\hline\hline
 & & & & \\
\end{tabular}

References \newline
1.  CRB   \newline
2.  Hummel (1980) (Westerbork Synthesis Radio Telescope) \newline
3.  Vila et al. (1990) \newline
4.  Crane et al. (1992); Crane et al. (1993)  \newline
5.  Turner \& Ho (1994) \newline

\end{table}

\clearpage
\begin{table}

\begin{tabular}{ccccc} 
\multicolumn{5}{c}{TABLE 5. Soft X-ray \& 6 cm Radio Flux Ratios for Selected LINERs} \vspace{0.2cm} \\ \hline\hline
\multicolumn{1}{c}{Source} &\multicolumn{1}{c}{Radio Flux$^{a}$}
&\multicolumn{1}{c}{X-ray Flux$^{a}$}
&\multicolumn{1}{c}{log [X-Ray/Radio]} &\multicolumn{1}{c}{Ref}\\

& $10^{-27}$  & 10$^{-30}$ & & \\ \hline

NGC 7331 & 0.121 & 1.93 --- 5.90 & ($-1.80$) --- ($-1.31$) & this Letter \& 1\\ 
NGC 4258 & 2.4   & 7.4 & $-2.1$ & 2 \& 3\\
M51	 & 1.6   & 3.2 & $-2.7$ & 5 \& 6\\
NGC 4736 & 3.6   & 4.1 & $-2.9$ & 3 \& 4\\
NGC 4736 & 6.1   & 4.1 & $-3.2$ & 3 \& 4\\
NGC 4579 & 40    & 5.3 & $-3.9$ & 3 \& 6\\
M81	 & 92    & 2.3 & $-4.6$ & 5 \& 6\\

\hline
\end{tabular}\newline

a.  erg s$^{-1}$ cm$^{-2}$ Hz$^{-1}$

References \newline
1.  CRB \newline
2.  Pietsch et al. (1994) \newline
3.  Hummel (1980) \newline
4.  Cui et al. (1997)    \newline
5.  Turner \& Ho (1994) \newline
6.  Serlemitsos et al. (1996) \newline  

\end{table}

\clearpage
\pagestyle{empty}

\figcaption{FIG. 1  A grey scale image of NGC 7331 and the field.  NGC 7331 is the circled object labeled \#1 and NGC 7335 is the circled object labeled \#2.  The X-ray detection of the nuclear source in NGC 7331 is within $5\fasec 0$ of prior radio observations of the nucleus given in Table 3.  The remaining eight
unidentified sources listed in Table 1 (\#3 - \#10) are indicated on this figure.
\label{fig1} }


\clearpage

\begin{thebibliography}{}
\bibitem[Afanasiev, Silchenko, \& Zasov 1989]{afa89}
Afanasiev, V. I., Silchenko, O. K., \& Zasov, A. V. 1989, A\&A, 213, L9
\bibitem[Argyle \& Clements 1990]{arg90}
Argyle, R. W. \& Clements E. D. 1990, Observatory, 110, 93
\bibitem[Bower 1992]{bow92}
Bower, G. A. 1992, PhD Thesis, Univ. Michigan
\bibitem[Bower et al. 1993]{bow93}
Bower, G. A., Richstone, D. O., Bothum, G. D., \& Heckman, T. M. 1993, ApJ, 402, 76
\bibitem[Burstein \& Heiles 1984]{bur84}
Burstein, D. \& Heiles C. 1984, ApJS, 54, 33
\bibitem[Ciardullo et al. 1988]{cia88}
Ciardullo, R., Rubin, V., Jacoby, G. H., Ford, H. C., \& Ford W. K. 1988, AJ 95, 438
\bibitem[Condon 1996]{con96}
Condon, J. J. 1996, ASP Conf, 103, 132. 
\bibitem[Cowan, Romanishin, \& Branch 1994]{cow94}
Cowan, J.J., Romanishin, W., \& Branch D. 1994, ApJ, 436, L139 (CRB)
\bibitem[Crane et al. 1993]{cra93}
Crane, P. C., Cowan, J. J., Dickel, J. R., \& Roberts, D. A. 1993, ApJ, 417, L61
\bibitem[Crane, Dickel, \& Cowan 1992]{cra92}
Crane, P. C., Dickel, J. R., \& Cowan, J. J. 1992, ApJ, 390, L9
\bibitem[Cui et al. 1997]{cui97}
Cui, W., Feldkhun, D., \& Braun, R. 1997, ApJ, 477, 693
\bibitem[Dahlem et al. 1995]{dah95}
Dahlem, M., Heckman, T. M., \& Fabbiano, G. 1995, ApJ, 442, L49
\bibitem[de Vaucouleurs et al. 1991]{dev91}
de Vaucouleurs, G., de Vaucouleurs, A, Corwin, H. G., Buta, R. J., Paturel, G., Fouqu{\'e}, P. 1991, Third Reference Catalog of Bright Galaxies  Springer-Verlag, New York (RC3)
\bibitem[Fabbiano 1996]{fab96}
Fabbiano, G. 1996, ASP Conf, 103, 56
\bibitem[Heckman 1996]{hec96}
Heckman, T. 1996, ASP Conf, 103, 241
\bibitem[Ho et al. 1996]{ho96}
Ho, L. C., Filippenko, A. V., \& Sargent W. L. W. 1996, ApJ 462, 18.
\bibitem[Ho 1998]{ho98}
Ho, L. C. 1998, Observational Evidence for Black Holes in the Universe, in print
\bibitem[Hughes et al. 1998]{hug98}
Hughes, S. M. G., Han, M., Hoessel, J., Freeman, W. L., Kennicut, R. C., Mould, J. R., Saha, A., Stetson, P. B.,  Madore, B. F., Silbermann, N. A., Harding, P., Ferrarese, L., Ford, H., Gibson, B. K., Graham, J. A., Hill, R., Huchra, J., Illingworth, G. D., Phelps, R., \& Sakai, S. 1998 ApJ, 501, 32
\bibitem[Hummel 1980]{hum80}
Hummel, E. 1980, A\&AS, 41, 151
\bibitem[Keel 1983]{kee83}
Keel, W. C. 1983, ApJ, 269, 466
\bibitem[Klemola et al. 1987]{kle87}
Klemola, A. R., Jones, B. F., \& Hanson, R. B., 1987, AJ, 94, 501
\bibitem[Koratkar et al. 1995]{kor95a}
Koratkar, A., Deustus. S. E., Heckman, T., Filippenko, A. V., Ho, L. C., \& Rao, M., 1995, ApJ, 440, 132
\bibitem[Kormendy \& Richstone 1995]{kor95b}
Kormendy, J. \& Richstone, D. 1995, ARA\&A, 33, 581
\bibitem[Long et al. 1996]{lon96}
Long, K. S., Charles, P. A., Blair, W. P., \& Gordon, S. M., 1996,
ApJ, 466, 750
\bibitem[Mahadevan 1997]{mah97}
Mahadevan, R. 1997, ApJ, 477, 585
\bibitem[Mediavilla et al. 1997]{med97}
Mediavilla, E., Arribas, S., Garc\'ia-Lorenzo, B., \& del Brugo, C. 1997, ApJ, 488, 682
\bibitem[Melia 1992]{mel92}
Melia. F. 1992, ApJ, 398, L95
\bibitem[Nandra et al. 1997]{nan97}
Nandra, K., George, I. M., Mushotzky, R. F., Turner, T. J., \& Yaqoob, T. 1997, ApJ, 477, 602
\bibitem[Narayan \& Yi 1995a]{nar95}
Narayan, R. \& Yi, I. 1995, ApJ, 444, 231
\bibitem[Pietsch et al. 1994]{pie94}
Pietsch, W., Volger, A., Jain, A., \& Klein, U. 1994, A\&A, 284, 386
\bibitem[ROSAT Users Guide 1994]{ros94}
ROSAT Users Guide, Max-Planck-Institut, 1994
\bibitem[Rubin et al. 1965]{rub65}
Rubin, V. C., Burbidge, E. M., Burbidge, G. R., \& Crampin, D. J. 1965, ApJ, 141, 759
\bibitem[Serlemitsos et al. 1996]{ser96}
Serlemitsos, P., Ptak, A., \& Yaqoob, T. 1996, ASP Conf., 103, 70
\bibitem[Smith 1998]{smi98}
Smith, B. J. 1998, ApJ, 500, 181
\bibitem[Stark et al. 1992]{sta92}
Stark, A. A., Gammie, C. F., Wilson, R. W., Bally, J., Linke, R. A., Heiles, C.,
\& Hurwitz, M. 1992, ApJS, 79, 77
\bibitem[Tosaki \& Shioya 1997]{tos97}
Tosaki, T. \& Shioya, Y. 1997, ApJ, 484, 664
\bibitem[Trotter et al. 1998]{tro98}
Trotter, A. S., Greenhill, L., J., Moran, J. M., Reid, M. J., Irwin, J. A. \& Lo, K. 1998, ApJ, 495, 740
\bibitem[Turner \& Ho 1994]{tur94}
Turner, J. L., \& Ho, P. T. P. 1994, ApJ, 421, 122
\bibitem[van Steenberg \& Shull 1988]{van88}
van Steenberg, M. E. \& Shull, J. M. 1988, A Decade of UV Astronomy with the IUE
Satellite, Vol. 2 (Noordwijk, The Netherlands: ESTEC), 219
\bibitem[Vila et al. 1990]{vil90}
Vila, M. B., Pedlar, A., Davies, R. D., Hummel, E., \& Axon, D. J. 1990, MNRAS, 242, 379
\bibitem[Young \& Scoville 1982]{you82}
Young, J. S., \& Scoville, N. 1982, ApJ, 260, L41

\end{thebibliography}
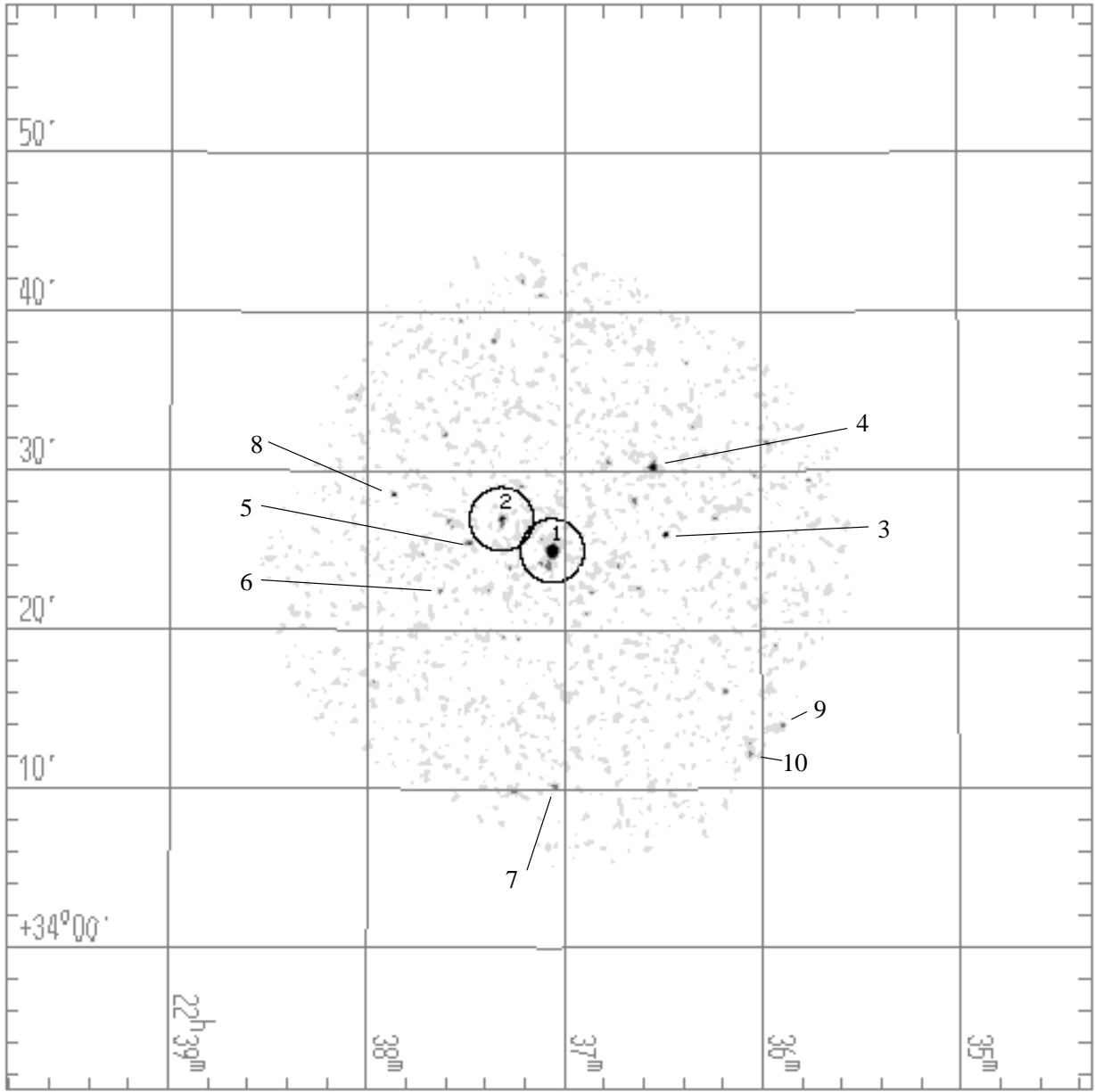
\end{document}